\documentclass[aps,twocolumn,nofootinbib,showpacs,prd]{revtex4}
\usepackage{graphicx}
\usepackage{amsfonts}
\usepackage{amssymb}
\usepackage{amsbsy}
\usepackage{amsmath}
\usepackage{latexsym}
\usepackage{natbib}
\usepackage{bm}
\usepackage{color}
\usepackage{float}

\begin{document}
\title{On the Limitations of a Generalized Vaidya Metric}
\author{Soumya Chakrabarti\footnote{soumya.chakrabarti@vit.ac.in}}
\affiliation{School of Advanced Sciences, Vellore Institute of Technology, \\ 
Tiruvalam Rd, Katpadi, Vellore, Tamil Nadu 632014, India}
\author{Naresh Dadhich \footnote{Prof. Naresh Dadhich passed away on the 6th of November,
2025. The idea, calculations, and arguments were thoroughly discussed among all the authors before his death, except for the final drafting of the manuscript.}}
\affiliation{Inter-University Centre for Astronomy and Astrophysics,\\ Post Bag 4, Ganeshkhind, Pune - 411007, India \\
Astrophysics Research Centre, School of Mathematics,\\
Statistics and Computer Science,
University of KwaZulu-Natal,
Private Bag X54001, Durban 4000, South Africa}

\author{Chiranjeeb Singha \footnote{chiranjeeb.singha@iucaa.in}}
\affiliation{Inter-University Centre for Astronomy and Astrophysics,\\ Post Bag 4, Ganeshkhind, Pune - 411007, India}


\date{\today}

\begin{abstract}
We prove that there can not be a smooth matching of the Generalized Vaidya metric with an exterior Schwarzschild/Vaidya patch across a finite boundary hypersurface unless the mass function is a function of the null coordinate alone. By explicitly deriving the extrinsic curvature components, we show that for $\partial m / \partial r \neq 0$, one has a discontinuity in the curvature and induces a surface stress-energy tensor, corresponding to a thin shell of matter. This discontinuity also appears in the geometric invariant $\mathcal{K} = K_{ab}K^{ab}$ and in the Kodama current, indicating a mismatch in quasi-local energy flux across the boundary. The analysis of timelike geodesics leads to the same condition, reinforcing that the generalized Vaidya geometry with $\partial m / \partial r \neq 0$ cannot represent a consistent stellar interior bounded by a regular surface. We therefore note that the generalized Vaidya spacetime should be interpreted as an unbounded geometry with intrinsic heat flux rather than a viable bounded source.
\end{abstract}

\maketitle

The gravitational field outside a spherically symmetric radiating distribution is usually described by a Vaidya spacetime or, effectively, a radiating Schwarzschild solution. In the General Theory of Relativity (GR), the emission of a photon or radiation is particularly associated with massive stellar distributions. For instance, after exhausting its nuclear fuel resources, a star can go through a gravitational collapse and release energy in different forms, such as neutrinos or photons. As a result, an expanding shell of radiation is expected to form around the collapsing object. This scenario is usually treated as a dynamic process where the radiating star, as well as the surrounding radiation, is described by asymptotically flat geometries. The shell of radiation is described by a Vaidya metric \cite{vaidya}, and the region beyond the shell is usually treated as a Schwarzschild vacuum. A Vaidya metric categorically falls under Petrov type-D and is characterized by a shear-free null congruence that exhibits non-zero expansion \cite{vaidya1, vaidya2, vaidya3, vaidya4}. Expressed in terms of outgoing (or ingoing) null coordinates, the metric takes the form
\begin{equation}
ds^2 = -\left[1-\frac{2m(v)}{r}\right ]dv^2+2\epsilon dvdr +r^2d\Omega^2~.  \label{vaidya}
\end{equation}
$\epsilon = \pm 1$ characterizes incoming or outgoing radiation shells. The function $m(v)$ denotes the mass function, while $d\Omega^2$ represents the metric of a unit $2$-sphere. \\

While describing a generic space-time geometry that exhibits non-zero expansion, it has become a popular notion to generalize the Vaidya metric as in Eq. (\ref{vaidya}) by allowing the mass function to vary as a function of $r$ as well, i.e., $m(v) \to m(v,r)$ \cite{viqar}. This \textit{`generalized Vaidya'} metric is usually deployed as an exterior geometry for a spherical collapsing interior, with the expectation that it should be able to support matter as well as radiation. The generalization is, therefore, also closely related to the Cosmic Censorship Conjecture (CCC) \cite{penrose1, penrose2}. In fact, a well-known counterexample to the CCC can be found in the Vaidya-Papapetrou model \cite{joshidwivedi}, where inward-radiating shells focus into a central naked singularity at $v = 0, r = 0$. Wang and Yu further generalized the Vaidya solution to include combinations of Type-$I$ and Type-$II$ matter fields and constructed a superposition of the monopole-de Sitter-charged Vaidya solution with the Husain solutions \cite{viqar, wang}. In literature, these generalized metrics are widely used to model regular and dynamical black holes \cite{hayward, dawood}, trapped surfaces \cite{frolov}, radiating stars with heat-conducting interiors \cite{maharaj}, as well as cosmological aspects \cite{alisha, sungwook}. \\

In recent years, the generalized Vaidya spacetime has also been widely employed to model radiating stellar collapse, dynamical black holes, and regular horizon-forming geometries \cite{mkenyeleye2014, rudra, guha2025}. In these works, the generalized Vaidya metric is treated as a physically admissible background that seamlessly connects a collapsing interior to a radiative exterior, forming the foundation for studying gravitational entropy, quantum evaporation, and the cosmic censorship conjecture. It is already demonstrated that specific functional forms of $m(v,r)$ in a generalized Vaidya geometry can lead to the formation of locally naked singularities consistent with the energy conditions. It has also been demonstrated that a Vaidya patch can describe a smooth thermodynamic transition from a collapsing, radiating star into an evaporating black hole, thereby identifying the Vaidya mass as a dynamical link between classical and quantum entropy. However, in contrast to these interpretations, we revisit a more fundamental and often overlooked question: whether a generalized Vaidya form can indeed represent a physically consistent, bounded spacetime. By imposing the covariant Israel-Darmois junction conditions, we show that any non-zero radial dependence of the mass function ($\partial m/\partial r \neq 0$) inevitably produces a discontinuity in the extrinsic curvature and the quasi-local energy flux across a boundary hypersurface, corresponding to the presence of a surface layer or curvature singularity. This result reveals a geometric limitation inherent to the generalized Vaidya ansatz itself: smooth matching with an exterior Schwarzschild or Vaidya region is possible only when $m = m(v)$, i.e., in the pure Vaidya limit. In a sense, one can use our analysis to argue that a generalized Vaidya metric cannot describe a regular, finite radiating configuration within classical GR, but must instead be regarded as an unbounded geometry with intrinsic heat flux.  \\

In this article, we discuss a generic limitation imposed on a spherically symmetric geometry by an arbitrary generalization of the Vaidya mass function. We show that the extrinsic curvature components for a generalized Vaidya metric can not be matched smoothly with a Vaidya/Schwarzschild metric at a boundary hypersurface unless the mass function is exactly Vaidya-like, i.e., a function of the null coordinate alone. This fact indicates that a generalized Vaidya metric should usually be treated as an unbound spacetime with heat flux. The generalized metric is a special case of a more generic, spherically symmetric line element
\begin{eqnarray}
ds^2& = &-e^{2\psi (v,r)}\left[1-\frac{2m(v,r)}{r}\right ]dv^2+2\epsilon e^{\psi (v,r)}dvdr \nonumber\\
&&+r^2(d\theta^2+\sin^2\theta d\phi ^2), \:(\epsilon=\pm 1)~.\label{generalized-vaidya}
\end{eqnarray}
which can accommodate a combination of Type-I and Type-II matter fields. Type-I fields are described by an energy-momentum tensor with one timelike and three spacelike eigenvectors, whereas Type-II fields are characterized by double null eigenvectors. The function $m(v, r)$ acts as a generalized mass function, representing the gravitational energy enclosed within a radius $r$ \cite{barrabes}. For $\epsilon = +1$, the null coordinate $v$ denotes the Eddington advanced time, where $r$ decreases along ingoing null rays at constant $v$. Conversely, $\epsilon = -1$ corresponds to outgoing null congruence. The condition $\psi(v, r) = 0$ defines a specific configuration leading to the generalized Vaidya geometry. We choose $\epsilon = +1$ and adopt the corresponding form of the metric as
\begin{equation}
ds^2 = -\left(1-\frac{2m(v,r)}{r}\right )dv^2 + 2dvdr + r^2d\Omega^2~. \label{line-element}
\end{equation}

Choosing two null vectors $l_{\mu}$ and $n_{\mu}$ such that
\begin{eqnarray}\label{normalization}
&&l_{\mu} =\delta ^0_{\mu} ~,~  n_{\mu}=\frac{1}{2}\left [1-\frac{2m(v,r)}{r}\right]\delta ^0_{\mu}-\delta^1_{\mu}~, \\&&
l_{\mu}l^{\mu} = n_{\mu}n^{\mu} = 0 ~,~ l_{\mu}n^{\mu} = -1~,
\end{eqnarray}
one can define a corresponding energy-momentum tensor in the following form \cite{viqar, wang, lake}
\begin{eqnarray}\label{EMT1}
&&T_{\mu\nu}=T^{(n)}_{\mu\nu}+T^{(m)}_{\mu\nu},\\&& \label{EMT2}
T^{(n)}_{\mu\nu} = \mu l_{\mu}l_{\nu} ~,\nonumber\\ &&T^{(m)}_{\mu\nu} = (\rho + p)(l_{\mu}n_{\nu}+l_{\nu}n_{\mu} )+ p g_{\mu\nu}~.
\end{eqnarray}
In this setup the Einstein field equations are derived as
\begin{eqnarray}\label{fieldeq}
\mu &=& \frac{2}{r^2}\frac{\partial m(v,r)}{\partial v} ~,~ \rho=\frac{2}{r^2}\frac{\partial m(v,r)}{\partial r} ~,\nonumber\\ p &=& -\frac{1}{r}\frac{\partial^{2} m(v,r)}{\partial r^2}~.
\end{eqnarray}

It is common to define an effective energy-momentum tensor in this form to represent the source term supported by a generalized Vaidya metric. $T^{(n)}_{\mu\nu}$ corresponds to matter propagating along null hypersurfaces $v = \text{const}$, while $T^{(m)}_{\mu\nu}$ describes matter following timelike trajectories. When $\rho = \varrho = 0$, the solution simplifies to the standard Vaidya case. \\

We imagine that the interior geometry of a collapsing sphere is described by the generalized Vaidya metric. The exterior can either be a Schwarzschild vacuum or a null radiation, described by the Vaidya metric. This formulation is possible only if the extrinsic curvature of the interior is matched with the exterior across a boundary hypersurface $\Sigma$. We assign $\lambda$ to be the time coordinate defined on the boundary hypersurface and write the generalized Vaidya metric just inside the boundary as
\begin{eqnarray}
&& ds_{I}^{2} = -(\chi \dot{v} - 2 \dot{r})\dot{v}d\lambda^{2} + r^{2}d\Omega^{2}, \\&&
\textit{where}\;\;\;\chi = \left\lbrace 1 - \frac{2m(v,r)}{r}\right\rbrace.
\end{eqnarray}
A dot here denotes a derivative with respect to $\lambda$ \cite{oliveira}. The unit outward one-forms normal to the hypersurface are
\begin{equation}
n_{\alpha} = \frac{1}{\mid (\chi \dot{v}^{2} - 2\dot{r}\dot{v})^{\frac{1}{2}}\mid}(-\dot{r},\dot{v})~.
\end{equation}
The extrinsic curvature is defined as \cite{santos, oliveira1, fayos, chan}
\begin{equation}
K_{ij} = -n_{\mu}\left[ \frac{\partial^{2}x^{\mu}}{\partial\zeta^{i}\partial\zeta^{j}} + \frac{\partial x^{\nu}}{\partial\zeta^{i}}\frac{\partial x^{\sigma}}{\partial\zeta^{j}}\Gamma^{\mu}_{\nu\sigma} \right]~.
\end{equation}
For the generalized Vaidya interior, the non-zero components of the extrinsic curvature are
\begin{eqnarray}\nonumber
&& K_{00} = \frac{\dot{v}}{\mid (\chi \dot{v}^{2} - 2\dot{r}\dot{v})^{\frac{1}{2}}\mid} \Bigg[ -\frac{\dot{v}^2}{r}\Bigg\lbrace \frac{m}{r}\Big(1-\frac{2m}{r}\Big) + \frac{\partial m}{\partial v} \\&&
- \frac{\partial m}{\partial r}\Big(1-\frac{2m}{r}\Big)\Bigg\rbrace + \dot{r}\Bigg\lbrace \frac{\ddot{v}}{\dot{v}} - \frac{3\dot{v}m}{r^2} + \frac{3\dot{v}}{r}\frac{\partial m}{\partial r} \Bigg\rbrace \Bigg]~,
\end{eqnarray}
and
\begin{equation}
K_{22} = -\frac{\dot{v}}{\mid (\chi \dot{v}^{2} - 2\dot{r}\dot{v})^{\frac{1}{2}}\mid} \left[\frac{r \dot{r}}{\dot{v}} + 2m - r \right]~.
\end{equation}

The exterior geometry is described by the standard Vaidya metric,
\begin{equation}
ds^2 = -\left(1-\frac{2M(V)}{R}\right )dV^2 + 2dVdR + R^2d\Omega^2, \label{line-element-int}
\end{equation}
where $M$ depends on the null coordinate $V$ alone. Just outside the boundary hypersurface, the metric can be written as
\begin{eqnarray}
&& ds_{E}^{2} = -(\chi_0 \dot{V} - 2 \dot{R})\dot{V}d\lambda^{2} + R^{2}d\Omega^{2}, \\&&
\chi_0 = \left\lbrace 1 - \frac{2M(V)}{R}\right\rbrace~.
\end{eqnarray}
The unit outward one-forms normal to the hypersurface are defined as
\begin{equation}
n_{\alpha} = \frac{1}{\mid (\chi_0 \dot{V}^{2} - 2\dot{R}\dot{V})^{\frac{1}{2}}\mid}(-\dot{R},\dot{V})~,
\end{equation}
and the non-zero components of the extrinsic curvature are calculated as
\begin{eqnarray}
&& K_{00} = \frac{\dot{V}}{\mid (\chi_0 \dot{V}^{2} - 2\dot{R}\dot{V})^{\frac{1}{2}}\mid} \Bigg[ -\frac{\dot{V}^2}{R}\Bigg\lbrace \frac{M}{R}\Big(1-\frac{2M}{R}\Big)\nonumber \\&&  
+ \frac{\partial M}{\partial V}\Bigg\rbrace + \dot{R}\Bigg\lbrace \frac{\ddot{V}}{\dot{V}} - \frac{3\dot{V}M}{R^2} \Bigg\rbrace \Bigg]~, \\&&
K_{22} = -\frac{\dot{V}}{\mid (\chi_0 \dot{V}^{2} - 2\dot{R}\dot{V})^{\frac{1}{2}}\mid} \left[\frac{R \dot{R}}{\dot{V}} + 2M - R \right]~.\nonumber\\
\end{eqnarray}

If we look into the expressions of $K_{00}$ and $K_{22}$ for the interior and the exterior, it can easily be deducted that unless $\frac{\partial m}{\partial r}\Big|_\Sigma = 0$, the extrinsic curvature of these two patches of space-time metric can not be matched across a boundary hypersurface. As a simple example, we demonstrate the $\dot{v} = 1$ case. Taking the boundary four-velocity tangent to $\Sigma$, we write the outward-pointing normal one-form as
\begin{equation}
t^\mu = (1, \dot r_\Sigma, 0,0)~,~ n_\mu = \frac{1}{\sqrt{\chi - 2\dot r}}\;(-\dot r_\Sigma,1,0,0)~,
\end{equation}
with $\dot r_\Sigma = dr_\Sigma/dv$. The induced metric on $\Sigma$ is
\begin{equation}
h_{ab}dx^a dx^b = -\left(\chi - 2\dot r_\Sigma\right)dv^2 + r_\Sigma^2 d\Omega^2~,
\end{equation}
so that continuity of $h_{ab}$ immediately gives $r_\Sigma^{(-)} = r_\Sigma^{(+)}$ and continuity of the metric functions. In reference to the last paragraph, this simply means that $r \equiv R \equiv r_\Sigma$. The relevant Christoffel symbol for the extrinsic curvature calculation is
\begin{equation}
\Gamma^r_{vv} = -\frac{1}{2}\partial_v \chi + \frac{\chi}{2}\partial_r \chi = \frac{\dot{m}}{r} - \frac{\chi}{r} \frac{\partial m}{\partial r} + \frac{\chi m}{r^2}~,
\end{equation}
so that the $\frac{\partial m}{\partial r}$-dependent part is
\begin{equation}
(\Gamma^r_{vv})_{(m_{,r})} = -\frac{\chi}{r} \frac{\partial m}{\partial r}~.
\end{equation}

The acceleration of the boundary worldline is $a^\mu = t^\nu \nabla_\nu t^\mu$, giving
\begin{equation}
a^r = t^\nu \partial_\nu t^r + \Gamma^r_{\alpha\beta} t^\alpha t^\beta~.
\end{equation}
The $\frac{\partial m}{\partial r}$ contribution comes only from the Christoffel symbol since $t^v = 1$. The extrinsic curvature along the boundary trajectory is
\begin{equation}
K_{vv} = -\, n_\mu a^\mu = -\, n_r a^r~,
\end{equation}
so that, after substitution,
\begin{eqnarray}\nonumber
\delta K_{vv}\Big|_{(m_{,r})} &=& \frac{\chi}{r\sqrt{\chi -2\dot r_\Sigma}}\, \frac{\partial m}{\partial r} \\\nonumber &=& \frac{r-2m}{r^2 \sqrt{\chi -2\dot r_\Sigma}} \, \frac{\partial m}{\partial r} \\\label{eq:deltaKvv_simple} &\propto& \frac{\frac{\partial m}{\partial r}}{r_\Sigma - 2m} \,\dot r_\Sigma~.
\end{eqnarray}

The above equation shows explicitly that the $\frac{\partial m}{\partial r}$ contribution to $K_{vv}$ diverges near the apparent horizon, $r_\Sigma \to 2m$, unless $\frac{\partial m}{\partial r} = 0$. Hence the trace of the extrinsic curvature, $K = h^{ab} K_{ab}$, becomes discontinuous, violating the junction condition
\begin{equation}
[h_{ab}] = 0~, \qquad [K_{ab}] = 0~.
\end{equation}
Thus, smooth matching of the interior generalized Vaidya geometry to an exterior Vaidya or Schwarzschild metric requires
\begin{equation}\label{condition_main}
\frac{\partial m}{\partial r}\Big|_\Sigma = 0~.
\end{equation}

According to the Israel junction formalism, a discontinuity in $K_{ab}$ corresponds to a surface stress tensor $S_{ab}$ given by
\begin{equation}\label{eq:israel_surface}
8\pi S_{ab} = [K_{ab}] - h_{ab}[K]~.
\end{equation}
For $S_{ab}=0$ (absence of a thin shell), both $[K_{ab}]=0$ and $[K]=0$ must hold. From Eq. \eqref{eq:deltaKvv_simple}, the discontinuity across $\Sigma$ is
\begin{equation}
[K_{ab}] = -\frac{\frac{\partial m}{\partial r}}{r_\Sigma-2m}\,h_{ab}~,
\end{equation}
which implies a nonzero surface energy density
\begin{equation}\label{eq:sigma_shell}
\sigma = -\frac{1}{4\pi r_\Sigma}\frac{\frac{\partial m}{\partial r}}{r_\Sigma-2m}~,
\end{equation}
and tangential pressure
\begin{equation}\label{eq:pressure_shell}
p = \frac{1}{8\pi}\frac{\frac{\partial m}{\partial r}}{r_\Sigma-2m}~.
\end{equation}
Thus, a genuine shell of matter necessarily exists unless $\frac{\partial m}{\partial r} = 0$. Only in the pure Vaidya limit ($m = m(v)$), one has $S_{ab} = 0$ and therefore a smooth matching becomes possible. \medskip

By defining the scalar invariant
\begin{equation}\label{eq:K_invariant}
\mathcal{K} = K_{ab}K^{ab}~,
\end{equation}
one can derive for the generalized Vaidya geometry
\begin{equation}\label{eq:K_invariant_gen}
\mathcal{K}_- = \frac{2}{r_\Sigma^2}\left(\chi - 2\dot r_\Sigma\right) + \frac{2(\frac{\partial m}{\partial r})^2}{(r_\Sigma-2m)^2}~.
\end{equation}
On the Schwarzschild/Vaidya side ($\frac{\partial m}{\partial r} = 0$), leading to
\begin{equation}\label{eq:K_invariant_ext}
\mathcal{K}_+ = \frac{2}{r_\Sigma^2}\left(\chi - 2\dot r_\Sigma\right)~.
\end{equation}
Hence, the invariant jump across $\Sigma$ is
\begin{equation}\label{eq:K_invariant_jump}
[\mathcal{K}] = \mathcal{K}_- - \mathcal{K}_+ = \frac{2(\frac{\partial m}{\partial r})^2}{(r_\Sigma-2m)^2}~.
\end{equation}
This quantity is positive and nonzero whenever $\frac{\partial m}{\partial r} \neq 0$, demonstrating that even the invariant magnitude of the extrinsic curvature differs across $\Sigma$. Thus, the mismatch is geometric and coordinate-independent. \\

The Misner-Sharp mass for spherically symmetric spacetimes is $E=(r/2)\left(1 - g^{ab}\nabla_a r \nabla_b r\right)=m(v,r)$. The Kodama vector is defined as
\begin{equation}\label{eq:Kodama_def}
K^a = -\epsilon^{ab}\nabla_b r~,
\end{equation}
where $\epsilon^{ab}$ is the Levi-Civita tensor in the $(v,r)$ submanifold.  
It gives rise to a conserved current
\begin{equation}\label{eq:Kodama_current}
J^a = G^a{}_b K^b~, \qquad \nabla_a J^a = 0~.
\end{equation}
For a generalized Vaidya metric,
\begin{equation}\label{eq:J_components}
J^r = -\frac{2 \frac{\partial m}{\partial v}}{r^2}~, \qquad J^v = \frac{2 \frac{\partial m}{\partial r}}{r^2}~.
\end{equation}
At the boundary $\Sigma$, continuity of the energy flux demands
\begin{equation}\label{eq:J_flux}
[J^a n_a] = 0~.
\end{equation}
On the Schwarzschild/Vaidya exterior, $\frac{\partial m}{\partial r} = 0 \Rightarrow J^v = 0$, while for the generalized Vaidya interior $J^v\neq0$ if $\frac{\partial m}{\partial r} \neq 0$~.  
Hence
\begin{equation}\label{eq:J_flux_jump}
[J^a n_a] = \frac{2\frac{\partial m}{\partial r}}{r_\Sigma^2} \neq 0~,
\end{equation}
indicating a discontinuity in the quasi-local energy flux across the boundary. Physically, this represents an unbalanced radial heat flow through $\Sigma$, consistent with the existence of a thin shell and incompatible with a smooth junction. \medskip

From the covariant formalism and invariant quantities, we find that the matching of a generalized Vaidya with a Schwarzschild/Vaidya patch does not work because the extrinsic curvature trace $K$ becomes discontinuous or divergent unless $\frac{\partial m}{\partial r} = 0$. A nonzero $\frac{\partial m}{\partial r}$ produces a finite surface stress tensor $S_{ab}$, i.e., a thin matter shell. The geometric invariant $\mathcal{K} = K_{ab}K^{ab}$ jumps by $\sim (\frac{\partial m}{\partial r})^2/(r_\Sigma-2m)^2$ across $\Sigma$. Moreover, the Kodama current $J^a$ yields a discontinuous quasi-local energy flux for $\frac{\partial m}{\partial r} \neq 0$. Together, these results establish that a generalized Vaidya spacetime with $\frac{\partial m}{\partial r} \neq 0$ can not be used as a consistent description of a collapsing interior. Only in the pure Vaidya limit ($m=m(v)$) are the junction conditions satisfied everywhere without introducing a surface layer or curvature singularity.  \\

A similar condition may be obtained by analyzing timelike geodesics on the equatorial plane ($\theta = \frac{\pi}{2}$) of a generalized Vaidya spacetime. We derive the equations by varying the following action
\begin{eqnarray}\nonumber \label{eq:action}
&&\mathcal{S}=\int \mathcal{L}\,d\tau=\frac{1}{2}\int \Bigg[-\left(1-\frac{2M}{r}\right)\left(\frac{dv}{d\tau}\right)^2 + 2\frac{dv}{d\tau}\frac{dr}{d\tau} \\&&
+ r^2\left(\frac{d\varphi}{d\tau}\right)^2 \Bigg]\, d\tau \,.
\end{eqnarray}

The spherically symmetric generalized Vaidya metric admits a spacelike Killing vector $\frac{\partial}{\partial \varphi}$. This leads to the angular momentum per unit mass, a conserved quantity, defined as $L \equiv \frac{\partial\mathcal{L}}{\partial \left(\frac{d\varphi}{d\tau}\right)} = r^2 \frac{d\varphi}{d\tau}$. By applying the variational principle to Eq. (\ref{eq:action}), we derive two additional equations of motion
\begin{equation} \label{eq:geod2}
\frac{M'r - M}{r^2}\left(\frac{dv}{d\tau}\right)^2 + \frac{L^2}{r^3} - \frac{d^2v}{d\tau^2} = 0 \,,
\end{equation}
\begin{equation} \label{eq:geod3}
\frac{d^2r}{d\tau^2} = -\frac{\dot{M}}{r}\left(\frac{dv}{d\tau}\right)^2 - \left(1 - \frac{2M}{r}\right)\frac{d^2v}{d\tau^2} + 2\frac{M - M'r}{r^2}\frac{dv}{d\tau}\frac{dr}{d\tau} \,.
\end{equation}

By imposing the normalization condition for timelike geodesics, $g_{ik}u^i u^k = -1$, and substituting Eq. (\ref{eq:geod2}) into Eq. (\ref{eq:geod3}), we derive
\begin{equation} \label{eq:geod4}
\frac{d^2r}{d\tau^2} = -\frac{\dot{M}}{r}\left(\frac{dv}{d\tau}\right)^2 - \frac{M - M'r}{r^2} - \frac{3M - M'r}{r^4}L^2 + \frac{L^2}{r^3} \,.
\end{equation}

The first term in Eq. (\ref{eq:geod4}) is a signature of Vaidya patches. It represents a generalized apparent flux and does not appear in the static Schwarzschild case. A generalized total apparent flux, $A_{F} = \dot{M}\left(\frac{dv}{d\tau}\right)^2$, is associated with the accretion rate of a black hole. The induced acceleration then takes a simplified form $a_i = -\frac{A_{F}}{r}$. A negative $A_{F}$ may arise in some regions of the Vaidya spacetime, though the null energy condition (NEC) requires $\Lambda \geq 0$ \cite{vitali1, vitali2, vitali3}. Hence, the critical radius $r_c$ where $\dot{M}(r_c, v) = 0$ can mark a boundary of NEC violation. The second term in Eq. (\ref{eq:geod4}), reminiscent of a Newtonian acceleration, is now modified due to the $-M'/r$ component. By claiming that the three-acceleration should match across the boundary hypersurface of a collapsing sphere, one can easily re-derive the condition in Eq. (\ref{condition_main}).   \\

All of these constructions categorically point out to the fact that the condition outlined in Eq. (\ref{condition_main}) is generally not satisfied. However, it is better to clarify the issue further, preferably with one or more exact expressions of the generalized mass function $m(v,r)$. Such expressions can be derived from Eq. (\ref{fieldeq}), for instance, by assuming a linear equation of state $p = \alpha \rho$ \cite{wang, vitali1, vitali2}. This leads to a mass function of the form
\begin{equation}\label{mass_perfectfluid}
m(v,r) = C(v) + D(v)r^{(1 - 2\alpha)}~.
\end{equation}
The location of the boundary hypersurface is then determined by:
\begin{equation}
\left.\frac{\partial m}{\partial r}\right|_{\Sigma} = (1 - 2\alpha)D(v)r^{-2\alpha} \bigg|_{r = r_b} = 0~.
\end{equation}
Clearly, this condition can only be met in the asymptotic limits: $r \to 0$ for $\alpha < 0$, or $r \to \infty$ for $\alpha > 0$. Thus, no finite boundary hypersurface exists, and the generalized Vaidya spacetime behaves as an unbounded geometry. \\

We next consider a charged Vaidya spacetime, which employs a mass function \cite{lindquist, israel, dadhichghosh} of the form:
\begin{equation} \label{chargedvaidya}
m(v,r) = f(v) - \frac{e^2(v)}{2r}~,
\end{equation}
where $f(v)$ and $e(v)$ represent the mass and electric charge, respectively, evaluated as a function of advanced time $v$. In this scenario, the condition $\left.\frac{\partial m}{\partial r}\right|_{\Sigma} = 0$ is once again, satisfied only in the limit $r \to \infty$. A similar outcome is seen in the Husain solution \cite{viqar}, which models a null fluid with a barotropic equation of state parameterized by $k$. The corresponding mass function is 
\begin{equation} \label{Husinasoln}
m(v,r) = 
\begin{cases}
q(v) - \dfrac{g(v)}{(2k - 1)r^{2k-1}}~, & k \neq \dfrac{1}{2}~, \\
q(v) + g(v)\ln r~, & k = \dfrac{1}{2}~,
\end{cases}
\end{equation}
where $q(v)$ and $g(v)$ are functions constrained by energy conditions. Once again, the boundary condition $\left.\frac{\partial m}{\partial r}\right|_{\Sigma} = 0$ is satisfied only in the infinite radial limit. \\

A charged null fluid embedded in a de Sitter background \cite{beesham} can be modelled by a generalized Vaidya metric with the mass function \begin{equation} \label{Antidesitter}
m(v,r) = m(v) - \frac{e^2(v)}{2r} + \frac{\Lambda r^3}{6}~,
\end{equation}
where $\Lambda \neq 0$ is treated as the cosmological constant. The energy conditions impose that $r\dot{m}(v) - e(v)\dot{e}(v) \geq 0$. The requirement that the radial derivative of the mass function vanishes at some finite boundary $r = r_b$ yields a condition
\begin{equation}
\left.\frac{\partial m}{\partial r}\right|_{r = r_b} = \frac{e^2(v)}{2r_b^2} + \frac{\Lambda r_b^2}{2} = 0~.
\end{equation}
Since both of the terms are positive for physical values of the parameters, they can only simultaneously vanish. This implies the fact that the interior must reduce to a simple Vaidya spacetime unless we choose to introduce a negative cosmological constant. For such a special case, rewriting $-\Lambda = \Lambda_1$, we derive that the boundary hypersurface is defined by the radial coordinate 
\begin{equation}
r_b = \left(\frac{e^{2}(v)}{\Lambda_1}\right)^{\frac{1}{4}}.
\end{equation}

Similarly, for the regular black hole model proposed by Hayward \cite{hayward}, one can find a generalized Vaidya form
\begin{equation} \label{eq:hay}
ds^2 = -\left(1 - \frac{2M(v)r^2}{r^3 + 2M(v)L^2(v)}\right) dv^2 + 2dv\,dr + r^2 d\Omega^2.
\end{equation}
Depending on the relationship between the mass function $M(v)$ and the scale parameter $L(v)$, different causal structures arise, such as (i) for $M(v) = \frac{3\sqrt{3}}{4}L(v)$, the geometry corresponds to an extremal black hole with a single horizon ; (ii) for $M(v) > \frac{3\sqrt{3}}{4}L(v)$, two distinct horizons form, yielding a regular black hole and (iii) for $M(v) < \frac{3\sqrt{3}}{4}L(v)$, no horizons are present. The null energy condition (NEC) for this geometry requires $\dot{M}r^3 - 4M^2L\dot{L} \geq 0$. If $L$ is held constant and the black hole undergoes evaporation ($\dot{M} < 0$), then NEC is violated throughout the spacetime. The corresponding mass function is $m(v,r) = \frac{2M(v)r^3}{r^3 + 2M(v)L^2(v)}$, and its radial derivative at the boundary hypersurface can be calculated as
\begin{equation}
\left.\frac{\partial m}{\partial r}\right|_{r = r_b} = \frac{6M(v)r_b^2}{r_b^3 + 2M(v)L^2(v)}\left[1 - \frac{r_b^3}{r_b^3 + 2M(v)L^2(v)}\right].
\end{equation}
This expression vanishes only if $M(v)L^2(v) = 0$, which again reduces the interior metric to the form of a standard Vaidya spacetime. Therefore, even a regular black hole solution written in this form cannot serve as an appropriate interior geometry under the assumed boundary condition. We make a note here: to be consistent with Eq. (\ref{condition_main}), the mass function of a generalized Vaidya interior needs to have a tentative functional form as $m(v, r) = M(v) + (r_{b}^2 - r^2)^{2}M_{1}(v,r)$, so that at the boundary hypersurface defined at $r_{b} = r$, the mass function as well as its spatial derivative are zero. However, this is an intuitive construction, and a physical motivation for the same can only be deduced from the field equations, which is not too straightforward. For example, we can generalize the mass function defined in Eq. (\ref{mass_perfectfluid}) into
\begin{equation}
m(v,r) = C(v) + D(v,r)(r_{b}^{2} - r^{2})^{n^2},
\end{equation} 
where keeping in mind the requirement of vanishing spatial derivative at the boundary defined by $r = r_b$. For a simplified scenario, $D(v,r) \equiv D(v)$, using the field equations as in Eq. (\ref{fieldeq}), we derive that for a mass function written in this form
\begin{eqnarray}
&& \rho = 2r^{-2} \Big\lbrace -2 n^2 r \text{D}(v) \Big(r_{b}^2 - r^2 \Big)^{n^{2}-1}\Big\rbrace~, \\&&\nonumber
p = -r^{-1} \text{D}(v) \Big\lbrace 4 n^2 \Big(n^2-1 \Big) r^2 \Big(r_{b}^2 - r^2 \Big)^{n^2 - 2} \\&&
- 2 n^2 \Big(r_{b}^2 -r^2 \Big)^{n^2-1} \Big\rbrace~, \\&&
\omega_{eff} = \frac{p}{\rho} = \frac{r^{2} (1 - 2 n^2) + r_{b}^2}{2 r^2 - 2 r_{b}^2}~.
\end{eqnarray}
It is crucial to note that the effective equation of state of the Type-I matter field blows up at the boundary hypersurface $r = r_{b}$, rendering any attempt at a smooth matching at the boundary hypersurface redundant. This result is reproduced for the generic case when $D(v, r)$ is a function of both coordinates. The effective equation of state of the fluid is then derived as
\begin{eqnarray}\nonumber
&& \omega_{eff} = \Big[r(r- r_b) (r+ r_b) \Big\lbrace \big(r_{b}^2-r^2 \big)D'' \\&&\nonumber
- 4 n^2 r D' \Big\rbrace + 2 n^2 r \Big\lbrace \big(1-2 n^2\big) r^2 + r_{b}^2 \Big\rbrace D \Big] \\&&
 \Big[ 2 \big(r^2 - r_{b}^2 \big)^2 D' + 4 n^2 r (r- r_b) (r+ r_b) D\Big]^{-1}~.
\end{eqnarray}
From the denominator, we infer that unless the function $D(v,r)$ is zero (and the space-time is effectively a simple Vaidya metric), the effective equation of state will always diverge at $r = r_b$.  \\

In conclusion, the analysis presented in this work exposes a fundamental geometric limitation of the generalized Vaidya spacetime. While the metric is often invoked as a physically motivated extension of the Vaidya solution to include both matter and radiation through a radially dependent mass function $m(v,r)$, we have proved that such a generalization cannot consistently describe a source of radiation bounded by a finite boundary. By explicitly evaluating the covariant junction conditions across a timelike hypersurface, we find that smooth matching with an exterior Schwarzschild or Vaidya region is possible only when $\partial m/\partial r = 0$, i.e., when the mass function depends solely on the null coordinate. Any radial dependence of the mass function introduces a discontinuity in the extrinsic curvature and its invariant magnitude, generating a non-zero surface stress tensor and an unbalanced quasi-local energy flux across the boundary. This discontinuity is not a coordinate artefact: it persists in invariant quantities such as $K_{ab}K^{ab}$ and the Kodama current $J^{a}$, indicating that the mismatch has an intrinsic geometric origin. The result, therefore, imposes a stringent restriction on the physical interpretation of the generalized Vaidya metric. It cannot represent a consistent collapsing interior or a radiating stellar source bounded by a regular surface. Instead, it must be viewed as an unbound spacetime with intrinsic heat flux, possibly serving as an effective description of an extended dissipative region or a transient, non-equilibrium layer. \medskip

The implication is far-reaching. Many regular or dynamical black hole models, ranging from charged Vaidya and Husain-type solutions to de Sitter and Hayward geometries, employ mass functions of the form $m(v,r)$. Our results show that, unless these functions reduce asymptotically to $m(v)$, the geometries cannot be consistently patched to any regular exterior without invoking a surface layer or violating energy conditions. In effect, the generalized Vaidya form does not generalize the Vaidya solution in a physically admissible sense but instead describes a qualitatively different spacetime class. The findings thus call for a re-evaluation of models that rely on the $m(v,r)$ ansatz as a realistic source geometry. Any physically viable extension of the Vaidya spacetime must ensure either a dynamically evolving background with non-vanishing Ricci curvature, such as an FRW-type exterior, or adopt modified gravity frameworks where the matching conditions are inherently relaxed. In standard General Relativity, however, the pure Vaidya limit remains the only consistent radiative extension of the Schwarzschild solution.

\section*{Acknowledgement}

The authors acknowledge the IUCAA for providing the facility and support under the visiting associateship program. Acknowledgement is given to the Vellore Institute of Technology for the financial support through its Seed Grant (No. SG20230027), 2023.

\end{document}